# Terahertz Multiparameter Sensor using Fiber-Interrogated Frequency Selective Surface

Martin Girard,[1] Maksim Skorobogatiy,[1,*]

[1]*Département de Génie Physique, École Polytechnique de Montréal, Montréal, H3C 3A7, Canada*
*Corresponding author: maksim.skorobogatiy@polymtl.ca*

We propose using THz frequency selective surfaces interrogated with THz subwavelength optical fibers as sensors for monitoring of the optical properties of thick films that are brought in contact with such surfaces. Changes in the test film thickness and absorption losses can be measured simultaneously by interpreting variations in the spectral resonant features in the fiber transmission and reflection spectra. Particularly, changes in the film thickness induce strong shifts in the fiber transmission peaks, while changes in the film absorption induce notable amplitude variations in the fiber reflection peaks. When applied to the problem of monitoring of thickness and humidity content in the paper sheets, the proposed system shows sensitivity of 1.5 GHz / 10 µm to changes in the paper thickness, and sensitivity of 0.6 / (% of water by volume) to changes in the paper humidity. We estimate the detection limit of our device to be 10 µm for the paper thickness variation and 0.02% (of water by volume) for the paper humidity variation. Proposed sensor is implemented in the spirit of integrated optics, where a point device based on the frequency selective surface is interrogated with a THz fiber that is used for remote delivery of THz radiation.

## 1. Introduction

Thickness control and monitoring of other physical properties is an essential necessity in film manufacturing. Rapid online characterization is well developed for thin transparent films (<1 µm) and it is usually based on optical interference or spectroscopic techniques [1]. However, for thick opaque films (10µm-1mm), rapid online techniques for multiparameter characterization are much less developed.

In particular, simultaneous characterization of paper thickness, density and humidity is an important industrial problem in paper manufacturing. Currently, one resorts to using several distinct sensors to do such a characterization, which require substantial integration engineering effort at the production line. Multiple sensors are typically integrated into scanning heads, which are then placed at various locations of the production line [2].

The paper thickness is usually measured by an electronic caliper, where two fingers touch the paper, and the distance between the fingers is determined by the electromagnetic measurement with a precision of 0.5 – 1 µm. This method, however, has significant disadvantages. As the paper sheet moves at high speeds (as high as 10m/s), the fingers can tear the paper. Moreover, the fingers can polish the paper, producing a glossy stripe along the length, while also suffering wear from the mechanical contact with the paper.

In the case of water content, the measurement is typically performed by interpreting transmission through paper of several distinct infrared wavelengths (usually 3 or 4) that are chosen to coincide with paper and water absorption lines. This is a non-contact characterization technique, which has a typical resolution 0.25% of water by volume.

As the paper industry desires new sensors for quality control, such as printability, fiber orientation, porosity, etc., the density of sensors in the production line will invariably increase. Multiplication of scanners is something that industry wants to avoid, both due to complexity of integration and maintenance, increased rates of failures, as well as due to added costs.

Recently, in [3] the authors reported using THz time domain spectroscopy (THz-TDS) to measure thickness and absorption of the paper films. This method profits from the relative transparency of paper to THz waves, and from the ability of THz-TDS setup to measure simultaneously the amplitude and phase of the electric field transmitted through the paper film under test. They experimentally measured the thickness with a 0.5 µm precision, while the moisture content measure had a precision of 0.25 % (per weight), with a confidence level of 68%. The sensor system presented in [3] is essentially a classic THz-TDS setup that uses bulky free space optics in a highly precise arrangement that can be challenging to maintain and service when installed on a real production line. The system can be prone to signal variations due to ambient humidity variations, unless the THz optics is sealed in the compartment with controllable atmosphere. A similar characterization technique was recently used in [4] with a goal of characterizing water content in plants. In that work, many of the engineering challenges mentioned above were identified and successfully addressed, and the authors managed to build a functional hand-held system. Finally, the use of terahertz time-of-flight measurement have been suggested in [5], where the authors demonstrated simultaneous measurement of the film refractive index and thickness with good precision, however, no absorption measurements were reported.

In this work we propose a novel concept for the multiparameter THz scanner, where the sensor is an integrated device that comprises a metamaterial-based transducer, which is interrogated with a THz optical fiber. In this arrangement, one can use either a broadband THz-TDS-type interrogation, or a simpler amplitude-based detection technique with narrow band THz sources. Importantly, the sensor is completely separated from the THz source, while THz optical fiber is used for power and signal delivery, as well as device interrogation. Sensing mechanism is based on detection of changes in the resonant features of a resonant structure that interacts with the test film. For the resonant transduction mechanism we use THz frequency selective surfaces (FSS) that have been extensively studied in the THz range due to ease of their fabrication. We have recently demonstrated in [6] that when a subwavelength fiber is coupled to a FSS, the transmission and reflection spectra of such a fiber show sharp (~1 GHz) Fano resonances. These resonances are a product of interaction between the fiber modes and the FSS substrate modes that are strongly coupled by the metallic split-ring resonators. Consequently, these resonances are highly sensitive to the structure of the FSS substrate. In this work, we propose using such resonances to measure simultaneously changes in the thickness and water content of a paper film placed underneath the FSS substrate (see Fig. 1). Changes in the paper thickness are monitored by registering spectral shifts of the peaks in the fiber transmission spectrum, while changes in the paper humidity are interpreted from variations in the peak amplitudes in the fiber reflection spectrum.

## 2. Device geometry

The proposed device consists of a subwavelength fiber of radius R = 200 µm suspended over a FSS constructed by patterning split-ring resonators (SRR) on a 320 µm dielectric slab. The geometry is illustrated in Fig. 1, with geometrical parameters h = 25 µm, Λ = 400 µm, r = 90 µm, w = 15 µm and θ = 30º. The fiber is made of low-loss plastic such as Polyethylene ($n_{fiber}$ = 1.55), while the slab is made of fused silica ($n_{substrate}$ = 1.97). The SRRs are assumed to be made of perfect electric conductors. The paper film under test is placed in direct contact with the FSS substrate.

Clausius-Mossotti model [7] along with liquid water permittivity measurements [8], as well as a double-Debye model [9] with Bruggeman effective medium [3] have been used to model dielectric constant of paper in the presence of humidity. In our simulations, we used a dispersionless permittivity for simplicity, based on results from [7, 8].

A Bloch-Floquet condition is used for band diagram calculations to find the optical states in the infinitely long fiber/FSS structure. For finite structures, the port boundary condition is used with S-parameter calculations in order to compute fiber transmission and reflection spectra. COMSOL commercial finite element software is used in all calculations. For port boundary conditions, we assume a single mode both at the input waveguide and at the output waveguide in the form of the fundamental $HE_{11}$ fiber mode, linearly polarized along the **z** axis. A periodic boundary condition is imposed in the **x** direction to simulate an infinite metamaterial. A few millimeters of air are present over and under the system in the **z** direction and the computational cell is terminated with a PML.

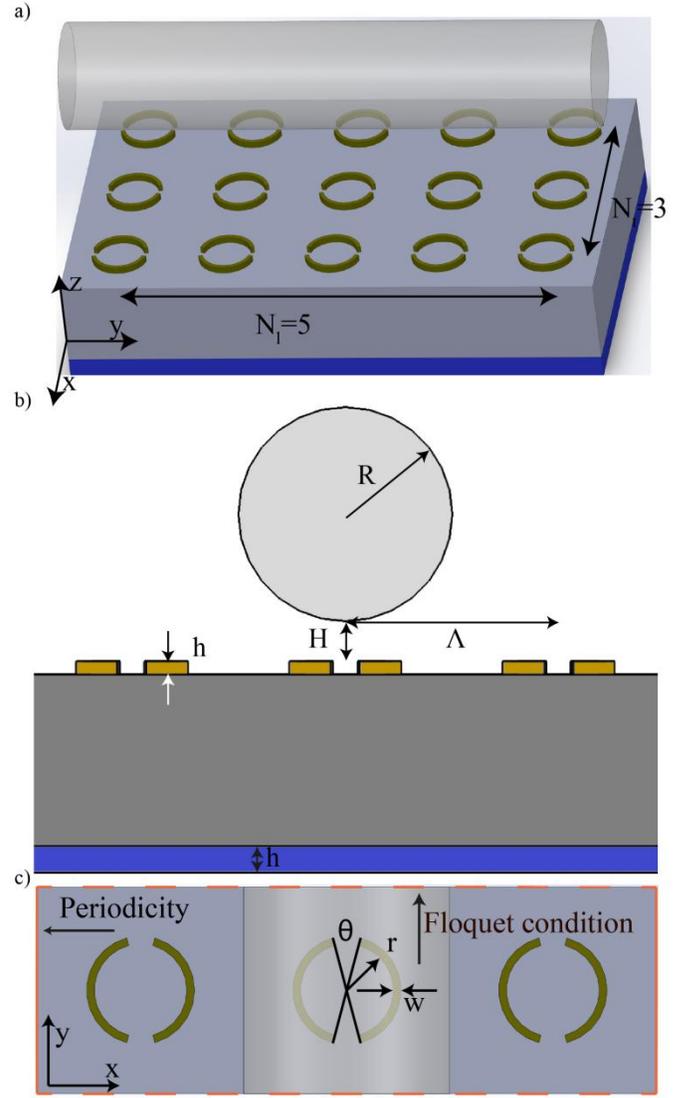

Fig. 1: a) 3D rendering of the frequency selective surface featuring 6 periods in the longitudinal direction and 3 periods in the transversal direction that is interrogated with a subwavelength fiber. Test film is placed in contact with the flat part of a substrate. b) Side view, d) top view of the unit cell that correspond to a) that we use in the band diagram calculations.

## 3. Fiber transmission and reflection spectra

In order to interpret the fiber transmission $t(f)$ and reflection $r(f)$ spectra, we use the scattering theory detailed in [10]. The scattering theory considers interaction of $N_r$ localized resonances with a single guided mode. In our case, the guided mode is the fundamental $HE_{11}$ fiber mode and the resonances are the FSS substrate modes coupled to the fiber mode via split-ring resonators. For a system with lossy resonator states, we use a transfer matrix method [11], which requires solving the equation:

$$\begin{pmatrix} t(f) \\ 0 \end{pmatrix} = M(f) \begin{pmatrix} 1 \\ r(f) \end{pmatrix} \qquad (1)$$

Where the transfer matrix $M(f)$ is defined by:

$$M(f) = \prod_{j=1}^{N_r} \begin{pmatrix} 1 - \dfrac{i\gamma_j}{f - f_j + i\Gamma_j} & -\dfrac{i\gamma_j}{f - f_j + i\Gamma_j} \\ \dfrac{i\gamma_j}{f - f_j + i\Gamma_j} & 1 + \dfrac{i\gamma_j}{f - f_j + i\Gamma_j} \end{pmatrix} \quad (2)$$

Where $f_j$, $\gamma_j$ and $\Gamma_j$ are the resonant frequency, coupling strength and losses that characterize the $j^{th}$ resonance. Particularly, resonant frequency corresponds to the phase matching point between fiber and FSS substrate modes, coupling parameter is characteristic of the spectral width of the resonance, while the loss parameter also includes additional absorption losses introduced by the lossy paper film, as well as radiation losses if the mode in question is leaky. In the rest of the paper we do not take into account absorption losses of the fiber material and FSS substrate, which is a valid approximation at lower frequencies (below 0.5 THz), where material losses of many dry dielectrics are small.

As we have mentioned earlier, variations in the paper thickness $h$ changes geometry and modal properties of the FSS cladding. Thus, we expect strong changes in the position of the fiber transmission resonances. In Fig. 2 we present transmission spectra $T = |t|^2$ trough a fiber suspended over a single period of a FSS ($N_l=1$, $N_t=3$, in Fig. 1) for various values of paper thicknesses $h = [50\ \mu m, 100\ \mu m, 150\ \mu m]$. In this simulation we assume lossless paper $\Gamma = 0$, while the real part of the paper permittivity is constant and equal to 2.25. As paper thickness is varied, clear spectral shifts are observed in the position of the transmission peaks. Peak parameters can be extracted from the transmission spectra by fitting them with lineforms defined by (1) and (2). The largest frequency shift is produced by the peak near 350 GHz. We note that the peak spectral shift and the peak width (coupling parameter) depends linearly on paper thickness (see Fig 2) in the studied parameter range. The spectral sensitivity, which is defined as a ratio of the frequency shift to the change in the layer thickness is calculated to be as high as 1.6 GHz / 10 µm.

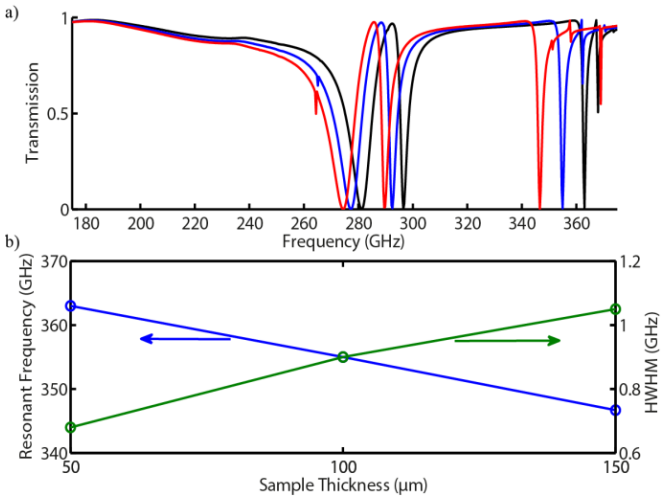

Fig. 2: a) Transmission through a subwavelength fiber coupled to a single cell of a FSS for various values of the paper thickness $t = 50\ \mu m$ (black), $100\ \mu m$ (blue), $150\ \mu m$ (red) b) For the peak near 350 GHz, dependence of the peak parameters on paper thickness.

We now study changes in the fiber reflection spectrum when non-zero paper absorption loss is considered. As follows from the lineshape (2), in the absence of losses, directly at the resonance, the reflection is perfect. When losses are introduced, the reflection coefficient at resonance decreases as:

$$R = |r|^2 = \frac{\gamma^2}{(\gamma + \Gamma)^2} \quad (3)$$

This lineshape is only valid for a single standing resonance, at it should be modified when two or more resonances are found in the direct vicinity of this resonance.

Losses $\Gamma$ in (3) can be varied by changing the imaginary part of the permittivity $\varepsilon''$ of paper. In Fig 3(a) we show reflection spectra for $\varepsilon''/\varepsilon_0 = [0.01, 0.03, 0.05]$. Assuming a Clausius-Mossotti model for humid paper along with an imaginary part value of the permittivity of water $\varepsilon''_w = 4.75$ [8], these values of the $\varepsilon''$ of paper correspond to the water content (by volume) of 0.85% to 4.1%.

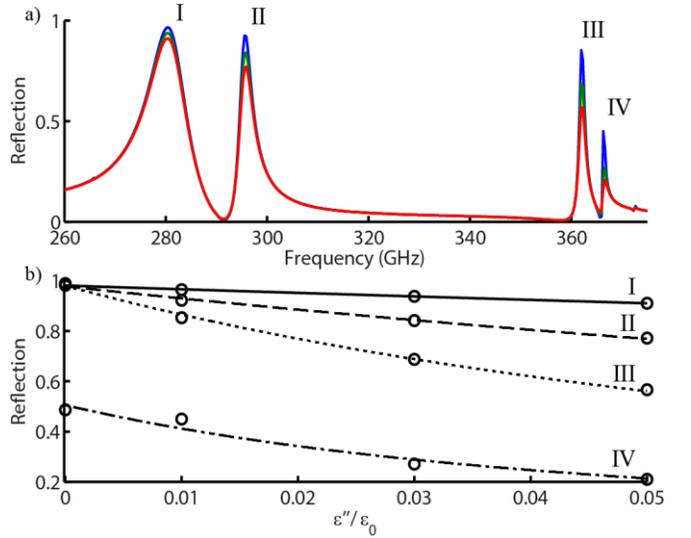

Fig. 3: a) Fiber reflection spectra for different values of the water content (by volume) w=0.85%, $\varepsilon''$=0.01 (blue); w=2.5%, $\varepsilon''$=0.03 (green); w=4.1%, $\varepsilon''$=0.05 (red). b) Value of the maximal reflection for four different resonances, I (solid), II (dashed), III (dotted), IV (dash-dotted), as a function of $\varepsilon''$ (paper thickness is fixed at 50 µm)

In Fig. 3(a) we present dependence of the reflection peaks (labeled in Fig 3(a) as I, II, III, IV) on paper losses. For each peak, lineshape (3) is used to fit the maximal value of the reflection as a function of the paper loss, where we assume that $\Gamma=C_\Gamma \cdot \varepsilon''$, where $C_\Gamma$ is a fitting constant. An excellent agreement is found between the numerical results (circles in Fig. 3(b)) and analytical fitting with (3) (continuous curves in Fig. 3(b)). Interestingly, when increasing paper losses, maximal value of the reflection coefficient falls much faster for narrower peaks (that are characterized by smaller values of the coupling parameter $\gamma$). This is well in accordance with the form of lineshape (3). Defining sensitivity to changes in the water content as a derivative of the value of the reflection coefficient with respect to the water concentration (by volume), we conclude that the sensitivity as high as 0.08 / (water % by volume) can be achieved at low humidity.

## 4. Band diagram versus S-parameter simulations

So far, we have presented transmission and reflection spectra of the subwavelength fiber coupled to a single cell of a FSS. In Fig. 4 we present transmission and reflection spectra for the subwavelength fiber coupled to a 10-period long section of a FSS. Compared to the case of a FSS with a single period, the spectra for a longer system features a much larger number of sharp resonances, typically with a Fano lineshape(for a detailed discussion of this phenomenon see our prior work [6]). This abundance of resonant peaks presents us with an opportunity of choosing the peak with optimal parameters for multiparameter detection. First of all, as we have mentioned earlier, using narrow resonances in reflection spectrum is advantageous for sensing changes in paper losses. At the same time, one cannot use peaks with high radiative loss to determine water content (see for example spectral dips in the vicinity of 300 GHz in Fig. 4) as the change in the reflection coefficient would be too small. Moreover, the peak width should be larger, while comparable, to the spectral resolution of a THz measurement setup. As an example, in [12] we have demonstrated experimentally using resonances of a THz fiber Bragg grating as narrow as 4 GHz, while using THz-TDS setup with 1.5 GHz resolution (600 ps – long scans).

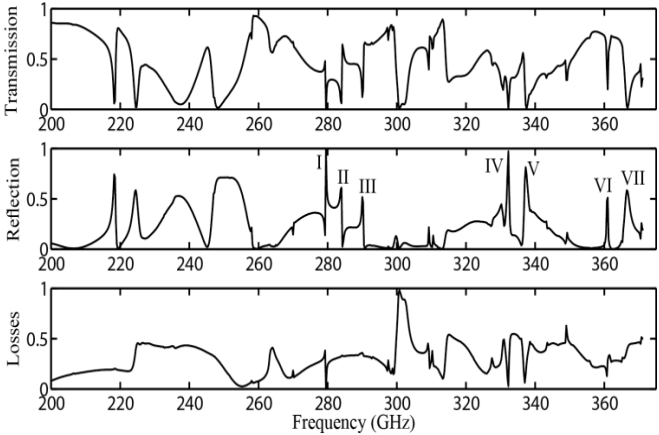

Fig. 4: Interrogation of a longer section of FSS (10 periods) with subwavelength fiber. a) transmission spectrum, b) reflection spectrum, and c) losses (1-R-T).

In what follows, we, therefore, focus on the peaks that feature low radiative losses and spectral widths in the range of 1 GHz. Seven particular resonances (labeled as I-VII) have been chosen in Fig. 4 for further studies..

To predict changes in the spectral position of resonant peaks caused by variations in the paper thickness, a direct approach would be to repeat the S-parameter calculations for each thickness value. This, however, is not very viable proposition if long sections of FSS are used. For example, in order to compute Fig. 4, it takes 80 GB of memory, and 5 days of time. Therefore, a more robust method is required. One such method uses fast band diagram calculations (also known as an ω(k) method) to predict positions of the resonant peaks in the fiber transmission spectra. As detailed in [6], resonances seen in Fig. 4 correspond to Van

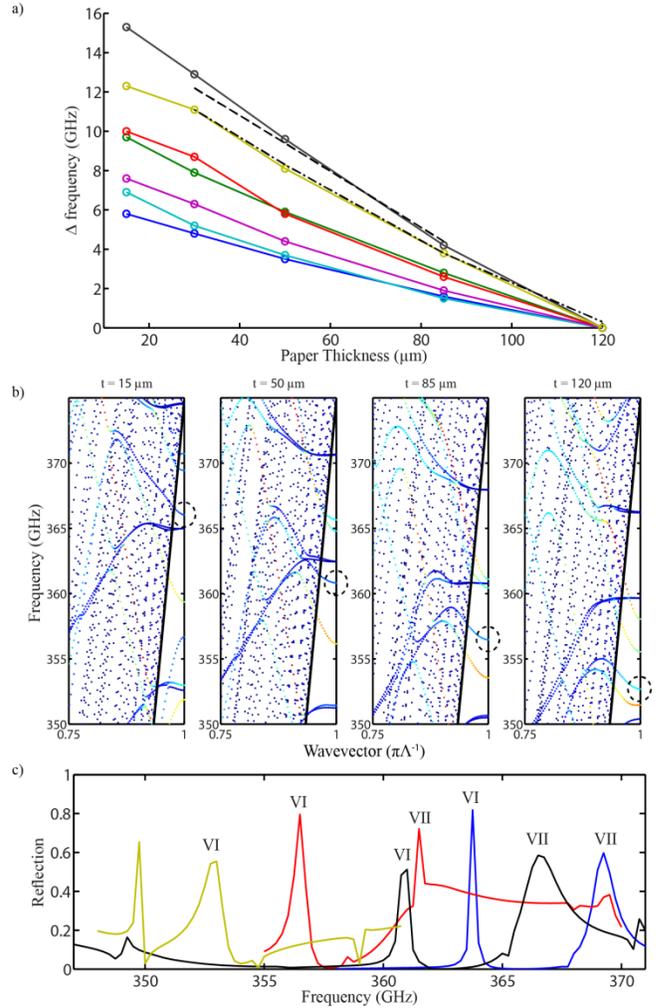

Fig. 5: a) Spectral change in the peak positions for different values of paper thickness. Presented data is for resonances I (blue), II (green), III (red), IV (teal), V (purple), VI (yellow), VII (black). In solid curves – results of the band diagram calculations, in dashed curves – results of the S-parameter calculation VI (dot-dashed) and VII (dashed). b) Band diagrams of the optical states for an infinite fiber / FSS system for various values of paper thicknesses of 15 μm, 50 μm 85 μm and 120 μm (resonance VI - dashed circles). c) S-parameter calculations for resonances VI and VII for paper thicknesses of 30 μm (blue), 50 μm (black), 85 μm (red) and 120 μm (yellow) used to confirm results of a) and b). A clear change of peak shape is seen for resonance VII between thicknesses of 50 μm and 85 μm. Resolution step for calculations is 0.25 GHz in a) and c)

Hove singularities in the band diagram of the optical states of an infinite fiber / FSS system. In Fig. 5(b) we demonstrate the band diagram method on the example of resonance VI. In this figure band diagrams are presented for various thicknesses of the paper layers and the positions of Van Hove singularities that correspond to resonance VI are highlighted in dashed circles. Spectral positions of the resonances can then be tracked simply by plotting a series of the band diagrams corresponding to the paper thickness of interest. This method is used to calculate solid curves in Fig. 5(b). One can also confirm the validity of this approach by performing much longer S-parameter calculations. Results of these simulations for resonances VI and VII are presented as dashed lines. An excellent correspondence between the band diagram approach and an S-parameter approach is observed. Finally, using data from Fig. 5(b), we can calculate spectral sensitivity of the peak position to changes in the paper thickness for various resonances to be 0.8 GHz / 10 μm, 1 GHz / 10 μm, 1.1 GHz / 10 μm, 1.1 GHz / 10 μm, 0.85 GHz / 10 μm, 1.5 GHz /10 μm and 1.6 GHz / 10 μm for resonances I-VII, respectively. It should be noted that for most resonances, shift in their spectral position is mostly linear with changes in the paper thickness (in the 0 – 85 μm paper thickness range). Notable exceptions are resonances III, IV and V. The reason for this is that the bands on which the Van Hove singularities are located move in frequency at a different rate than other adjacent bands. Due to interaction between these bands, they repulse each other in frequency and the displacement becomes non-linear.

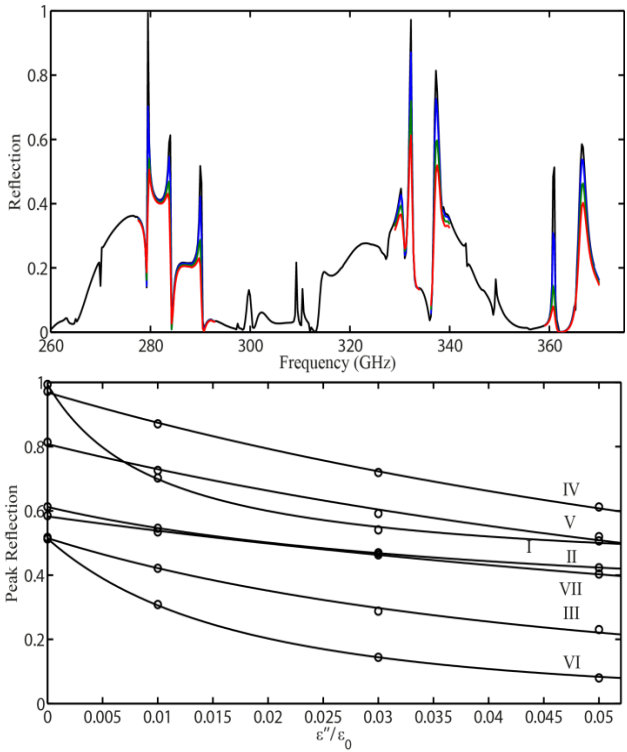

Fig. 6: a) Fiber reflection spectra calculated using S-parameter approach for different values of water content w = 0 (black), w = 0.85%, ε"/ε0 = 0.01 (blue); w = 2.5%, ε"/ε0 = 0.03 (green);w = 4.1%, ε"/ε0 = 0.05 (red) and a fixed paper thickness of 50 μm. Reduction in the maximal values of the reflection coefficient at resonances is clearly visible when the paper loss increases. b) Maximal value of the reflection coefficient for peaks I – VII. Circles – S-parameter calculations, solid curves - fitted lineshape (3).

Finally, we note that band diagram calculations (of the ω(k) type) cannot be used directly to compute resonance loss parameter $\Gamma$. Therefore, changes in the reflection coefficient have to be calculated using an S-parameter method. In Fig. 6(a) we present changes in the fiber reflection spectrum as a function of the paper loss (water content). As expected, higher paper losses result in smaller values of the reflection coefficient at resonance. To quantify this observation, in Fig. 6(b) we present maximal values of the reflection coefficient at various resonances (identified as I – VII in Fig. 4) as a function of the paper loss parameter ε"/ε0=0.01, 0.03, 0.05. These values of the imaginary part of the paper permittivity correspond to the 0.85%, 2.5% and 4.1% of water by volume in the paper. Individual data points (circles in Fig. 6(b)) are then fitted with lineform (3) (solid cirves in Fig. 6(b)). Excellent fit is observed with a simple lineshape (3) for all the resonances except I and II. For these two resonances it is necessary to add a constant baseline to the lineshape (3). This is because these two resonances are too closely spaced with respect to each other, therefore, a simple lineshape (3), which describes a single standing resonance, is no longer adequate. From the data in Fig. 6(b) we calculate that sensitivities as high as 52 / (ε"/ε0), or 0.60 / (% water by water) can be achieved at low humidity levels (peak I).If we assume that a change of 1% of the reflectivity is measurable, the limit of detection for water content is ~0.02%. This value can potentially be further decreased by using a thinner substrate, but one has to consider experimental problems related to very thin substrates and implementation of such a sensor in industry.

While more resonant peaks could be found at higher frequencies, material absorption of the FSS substrate and fiber material also increases with frequency. For example, absorption for fused silica changes from ~0.25 cm$^{-1}$ at 0.5 THz to ~1 cm$^{-1}$ at 0.75 THz to ~2 cm$^{-1}$ at 1 THz [13], while absorption loss of polyethylene varies from ~0.1 cm$^{-1}$ at 0.5 THz to ~0.3 cm$^{-1}$ at 1 THz [14]. As follows from the general considerations of the scattering theory (2), increase in the material losses will significantly reduce sensitivity of the resonances to changes in paper loss when resonance loss becomes comparable to the coupling strength parameter $\Gamma \sim \gamma$. Therefore, at higher frequencies, it is important to take into consideration material losses of the fiber / FSS system to produce a realistic analysis of sensor performance.

Another potential degrading factor is the uniformity of the fiber-FSS cell. Using a perfectly horizontal fiber is impossible in practice and the fiber-substrate distance, $H$, will change over the coupling length. Our previous study on this geometry has shown that the resonant frequency can be modeled using a simple perturbative model [6]:

$$f_r(H) = f_0 + \Delta f_r \cdot \exp(-H/H_f) \quad (4)$$

Where fr is the resonant frequency, f0 is the resonant frequency in the zero coupling limit, Δfr is the tuning range of the resonant and Hf is the characteristic fiber-metamaterial separation distance. Tuning range has been shown to be higher for low Q resonances and a typical value of Hf for such a resonance is in the hundreds of microns. Therefore, the system should not show high sensitivity to misalignment in fiber-FSS distances.

However, other potential non-uniformities may plague the system. Surface roughness of the fiber, substrate and paper layers can induce additional losses, lowering the sensitivity of the sensor. In the case where high losses (substrate absorption, scattering, etc.) are present, using a resonance showing nearly linear behavior on Fig. 6, such as resonance IV may prove a better choice. In that particular case, the maximum achievable LOD is of 0.14% of water content.

## 5. Conclusions

In conclusion, we have demonstrated a multiparameter sensor based on the frequency selective surface interrogated with a THz optical fiber that is capable of simultaneous measurement of both the paper thickness and water content. The transduction mechanism based on analysis of the fiber transmission and reflection spectra is analyzed using band diagram and S-parameter calculations. For the variations in the water content (paper humidity), sensitivities as high as 0.6 / (% of water by volume) are predicted. The corresponding detection limit of 0.02% of water variation by volume is estimated by assuming that reliable detection of 1% change in the value of the reflection coefficient is possible. For the variations in paper thickness, sensitivities as high as 1.5 GHz / 10 μm are predicted. The corresponding detection limit of 10 μm in paper thickness variation is estimated by assuming that spectral resolution of a TDS-THz setup is 1.5 GHz.

These values are superior to those previously reported for other types of metamaterial-based sensors of thickness [15, 16] for thick samples. These sensors used a resonance shift to measure thickness, with high sensitivity (up to 400 GHz / μm) up to ~20 μm thickness and nearly zero for thicker samples (~1-2 GHz shift in the 20 μm – 100 μm range). Moreover, our sensor exhibits a virtually linear response over a wide range of thicknesses, while sensors presented in [15, 16] show highly nonlinear exponential behavior. Although our sensor shows somewhat inferior sensitivity when compared to a direct THz-TDS-based thickness measurements detailed in [3], however, at the same time, it shows significantly higher sensitivities for the water content measurements. Furthermore, our approach does not require a coherent detection strategy, therefore, much cheaper incoherent narrow-band sources and detector can be used. Finally, the proposed sensor is implemented in the spirit of integrated optics, where a point device based on the FSS is interrogated with a THz fiber that is used for remote delivery of THz radiation. We believe that this paradigm is advantageous for building modular, highly sensitive, while practical THz sensors.